\documentclass{aa}
\usepackage{graphicx}
\usepackage{multirow}
\usepackage{txfonts}
\usepackage[authoryear]{natbib}       
\bibpunct{(}{)}{;}{a}{}{,}
\newcommand{\FeII}{[\ion{Fe}{II}]}

\begin{document}

\title{Light echo of V838 Monocerotis: \\
properties of the echoing medium
\thanks{Based on observations made with the NASA/ESA Hubble Space Telescope, 
obtained from the data archive at the Space Telescope Science Institute. 
STScI is operated by the Association of Universities for Research in Astronomy, 
Inc. under NASA contract NAS 5-26555.}}

\author{R. Tylenda\inst{1} \and T. Kami\'nski\inst{2}}  

\offprints{R. Tylenda}
\institute{Department for Astrophysics, N. Copernicus
            Astronomical Center, Rabia\'{n}ska 8,
            87-100 Toru\'{n}, Poland\\ 
            \email{tylenda@ncac.torun.pl} 
        \and Max-Planck Institut f\"ur Radioastronomie, Auf dem
            H\"ugel 69, 53121 Bonn, Germany}
\date{Received; accepted}
\abstract
{The light echo phenomenon that accompanied the 2002 eruption of
V838~Mon allows one to study the properties of the diffuse dusty matter in
the vicinity of the object.} 
{We are aiming at obtaining estimates of the optical thickness of the
circumstellar matter in front of V838~Mon, as well as optical properties of
dust grains in the echoing medium. In particular, we are interested in
studying whether the echoing medium can be responsible for the observed
faintness of the B-type companion of V838~Mon when compared to three B-type
stars that are seen in the vicinty of V838~Mon and are believed to be at 
the same distance as V838~Mon.}
{We used the V838~Mon light echo images obtained by the Hubble Space Telescope 
(HST) in different
filters and epochs. From the images we derived the total brightness of the echo
and its surface brightness. The results of the measurements were
compared to model light echoes. }
{The present study allowed us to estimate the optical
thickness of the matter in front of the object and 
the mean cosine value of the
scattering angle of dust grains in three HST filters.}
{ The optical thickness of the echoing matter is not sufficient to explain
the observed difference in brightness between the B-type companion of
V838~Mon and the other three B-type stars observed in the vicinity of V838~Mon.
Implications of this result are discussed. 
Our estimate of the mass of the diffuse matter seen in the light echo shows
that the matter cannot have resulted form a past mass loss activity of V838~Mon.
We probably observe remnants of an interstellar cloud from which V838~Mon and
other members of the observed cluster were formed.}

        
\titlerunning{Light echo of V838 Mon}
\authorrunning{Tylenda \& Kami\'nski }
\maketitle
%----------------------------------------------------------------------------------
\section{Introduction  \label{intro}}

V838 Monocerotis was discovered as an eruptive star at the beginning of 
January 2002. The main eruption started at the beginning of 
February 2002, however, and lasted about two months 
\citep[see e.g.][]{muna02,kimes02,crause03}. The main characteristic of the
eruption, which is different in V838~Mon from other stellar eruptions (e.g.
classical novae), was that the object evolved to progressively lower
effective temperatures and declined as a very late M-type supergiant
\citep[e.g.][]{tyl05}.

\citet{ts06} showed that the eruption of V838 Mon cannot be explained by
a thermonuclear runaway similar to classical novae, nor by a late He-shell
flash. They proposed, following \citet{st03}, that the event resulted from a
merger of two stars. This idea obtained strong support from an analysis of
archive observations of the progenitor of V1309 Sco. The
eruption of this object was observed in 2008 and was of the same type as that 
of V838 Mon \citep{mason10}. \citet{thk11}
showed that the event resulted from a merger of a contact binary.

The eruption of V838 Mon was accompanied by a spectacular light echo.
A series of fantastic images of the light echo have been
obtained by Hubble Space Telescope \citep[HST, see e.g.][]{bond03,bond07}. 
Analyses of the light
echo's structure and properties allowed astrophysicists to study the dusty medium 
around V838~Mon \citep[e.g.][]{tyl04} and to determine the distance to the object
\citep{sparks}.

V838 Mon has a spectroscopic companion, a B-type star
discovered by \citet{mdh02} in September--October 2002, 
when the eruption remnant became so cool that
it almost disappeared from the optical. The spectrum of the
companion, classified as B3\,V by \citet{mdh02}, 
was seen undisturbed until the fall of 2006, when an eclipse-like event
was observed \citep[e.g.][]{muna07}. The companion was eclipsed for about
two months.
In 2004/2005, emission lines of \FeII\ appeared in the
spectrum of V838~Mon. The lines strengthened with time and reached a
maximum at the time of the eclipse-like event. \citet{tks09} showed that the
profiles and evolution of the \FeII\ emission lines, as well as the
eclipse-like event resulted from interactions of the B-type star radiation with the
matter ejected by V838~Mon in the 2002 eruption.
After reappearance from the eclipse-like event, the companion
began to fade again and in 2008 traces of the B-type spectrum, as well as
of the \FeII\ emission lines, disappeared.
Apparently the companion became completely embedded in the dusty
ejecta of V838~Mon \citep{tks11}.

\citet{afbond} found a sparse, young cluster, containing three B-type
main-sequence stars, in the field of V838~Mon. The cluster is at a distance of
$\sim$6.2~kpc, which is very close to the value of $\sim$6.1~kpc derived by
\citet{sparks} for V838~Mon from their analysis of the polarization of the
light echo. The reddening of the cluster and that of V838~Mon and its
companion are also very similar ($E_{B-V} = 0.84$ versus 0.9, respectively).
It seems very likely that V838~Mon and its B-type companion are
also members of the cluster. 

However, there is a problem with the observed
brightness of the B-type companion of V838~Mon when compared to the
magnitudes of the B-type stars of the cluster. \citet{muna05} derived
$V$~=~16.05~$\pm$~0.05 for the companion of V838~Mon. 
Using the compilation of photometric data of 
V.~Goranskij\footnote{http://jet.sao.ru/jet/$\sim$goray/v838mon.htm},
one derives $V$~=16.19~$\pm$~0.03 from the data obtained in
September--November~2002.
The B-type companion is therefore fainter by 1.2--1.4~mag when compared to 
star~9 (classified as B3\,V) in the cluster of \citet{afbond}. 
\citet{afbond} interpreted this as
evidence that the companion is partly submerged in the dusty ejecta of
V838~Mon. However, if a dusty medium results
from a stellar eruption, it is very unlikely that it
is homogeneous. Therefore, if the ejecta flows in front of a star, one
expects to observe a high variability of the stellar light. 
This was not observed in the case of the B-type companion of V838~Mon
until the fall of 2006. As can be seen from the
light curve of V838~Mon between
fall~2002 and fall~2006 (see e.g. the data of Goranskij$^1$), 
the $U$ brightness (dominated by
the B-type companion in this epoch) remained constant within $\pm$0.1 mag.
The light curve exhibits the same behaviour in the $B$ band 
 after the 2002 eruption \citep[see][]{muna05}.
As discussed above, interactions of the V838~Mon ejecta with the companion
started at the earliest in the fall of 2006 and then indeed resulted in a high variability
of the observed light of the companion, eventually leading to the complete
disappearance of the star.

The problem of the observed brightness of the B-type companion remains
unresolved.
A possible explanation is that V838 Mon and its companion suffer from
additional, local extinction of $\Delta A_{V} \simeq 1.3$, compared to the three B-type 
stars of the cluster. A standard extinction of this value would imply 
additional reddening of $\Delta E_{B-V} \simeq 0.4$. This is significantly higher
than the observed difference between the reddening of the V838~Mon and its
companion, and that of the cluster ($\Delta E_{B-V}\la 0.1$, see above). 
This implies that the
local extinction, if present, would have to be almost grey in the
optical, i.e. $R = A_{V}/E_{B-V} \ga 10$, which would additionally 
imply local dust dominated by large grains ($a \ga 0.3~\mu$m).

This idea seems to be conceivable. Young stellar clusters are
usually known to contain a significant amount of diffuse matter consisting
of remnants of a more 
massive complex, from which the cluster was formed. For
V838~Mon there is observational evidence that diffuse matter is indeed present
in the near vicinity of the object. One piece of evidence is the light echo. Radio
observations in the CO rotational lines also show the presence of molecular matter,
most probably related to the dusty medium seen in the light echo
\citep{kmt,ktd}. The light echo offers possibilities to study
parameters and properties of the scattering matter. This is
the aim of the present study.

As we show in detail below, an analysis of the
brightness of the echo at different wavelengths and its evolution with time
allows us
to estimate several parameters of the scattering matter. This includes
the optical thickness of the matter, as well as
dependeces of the scattering coefficient on the wavelength and the
scattering angle. In this way we can verify  whether the local
extinction can be responsible for the relative faintness of the B-type companion of
V838~Mon when compared to the other B-type stars of the cluster.
The results of the echo study also allow us to derive constraints on the nature
of the dust grains and on the mass of the scattering material. The latter
is important for discussing the nature of the matter, i.e. whether it is
of interstellar origin or rather, as suggested by some authors
\citep[e.g.][]{bond03,bond07}, that it resulted from past mass loss 
activity of the progenitor of V838~Mon. This can have consequences for the
nature of the progenitor itself and the mechanism of the 2002 eruption.

%++++++++++++++++++++++++++++++++++++++++++++++++++++++++++++++++++++
\section{Observational material and its reduction\label{obs_sect}}

 We have used the archival images of V838~Mon's light echo obtained with
HST in 2002 on April~30, May~20, September~2, October~28, 
and December~17 (HST proposals 9587, 9588, and 9694). The first image was taken in 
the F435W filter only. The other ones were obtained in the F435W, F606W, and F814W filters. 
The observations were performed with the Advanced Camera for Surveys (ACS) combined 
with the detector of the Wide Field Channel (WCS). Some of the images were taken in 
the polarimetric mode, the rest in the standard (non-polarimetric) mode. 
The observations have been described in more detail in \cite{sparks}. 
The pipeline-reduced data, which we extracted from the HST archive, were further 
processed using the {\it multidrizzle}\footnote{http://stsdas.stsci.edu/multidrizzle} 
package (version 3.3.8). 
The non-polarimetric observations for a given date and filter were combined in 
{\it multidrizzle} to produce images in units of counts per second. 
The polarimetric observations were obtained with three retarder angles 
(0\degr, 60\degr, 120\degr). We first combined 
frames  with the same retarder angle
for a given filter and date. 
Then the  polarimetric images were scaled using calibration 
corrections taken from 
{\it ACS Data Handbook}\footnote{http://www.stsci.edu/hst/acs/documents/handbooks/currentDHB}. 
Finally the data were combined into the Stokes $I$ (total intensity) 
image in each filter. 

The following reduction was performed using 
Starlink\footnote{http://starlink.jach.hawaii.edu/starlink} packages, mainly 
with KAPPA and GAIA. We measured the background sky level and its variations in 
each image by performing statistics on a large number of pixels free of stellar 
and echo emission and instrumental defects. After correcting images for 
the sky level, those measurements were used to blank all pixels below 1.5$\sigma$ 
of sky variations and above the maximum echo brightness for a given 
date.
From the resulting images, 
we manually removed the remaining parts of stellar diffraction patterns and 
cosmic-ray hits. Finally, we measured the light echo region.

\section{Measurements  \label{resobs_sect}}

The results of the measurements are given in Table~\ref{measur_tab}. 
The upper, middle, and bottom parts of the table
present the results obtained with the F435W, F606W, and F814W filters, respectively. 
The dates of the observations are given in column~(1), while column~(2) notes 
the observing mode used to obtain the images. Total count rates from the echo 
are given in column~(3).
Column~(4) gives the total number of pixels, which registered a measurable
signal from the echo.
A total flux from the light echo can be calculated by multiplying  
the total count rate number by the {\it photflam} conversion factor
\citep{sirianni}, which is equal to
$3.08 \times 10^{-18}$, $7.725 \times 10^{-19}$, and $6.94 \times 10^{-19}$
 W\,m$^{-2}$\,$\mu$m$^{-1}$ per
count~s$^{-1}$ for the F435W, F606W, and F814W filters, respectively.
The resulted fluxes are presented in column~(5). The ACS pixel 
scale is $0\farcs05 \times 0\farcs05$, which allows one to easily calculate
a mean surface brightness, $S_{\rm B}$, from the mean count rate summed over 
400 pixels (corresponding to 1~arcsec$^2$) and the 
{\it photflam}. The results are listed in column~(6).

The data in columns~(3)--(6) refer only to the part of 
the light echo surface that
was not contaminated by field stars and V838~Mon itself.
The procedure of removing diffraction patterns of field stars 
introduced uncertainties in the measurements. 
This is particularly the case for the total flux,
which is expected to be systematically underestimated
in column~(5) of Table~\ref{measur_tab}. The effect is difficult to estimate, as
each time different parts of the echo were contaminated by stars. 
The mean surface brightness of the echo is expected to be less vulnerable 
to the procedure of cutting out the stellar patterns.

We have attempted to correct the measured total flux for this effect.
From the radius of the outer rim of the echo measured in \citet{tss05},
we calculated the total number of pixels expected inside the echo.
Then we defined a covering factor, $f_{\rm c}$, 
as a ratio of the number of pixels that
detected the echo (i.e. counted in column~4 of Table~\ref{measur_tab}), 
to the total number expected from the size of the
echo. This factor is generally between 0.7--1.0.
Assuming that $f_{\rm c}$ is $<$\,1.0 because of
the field stars' patterns and that the mean surface brightness is
statistically the same over the echo image, the corrected echo flux can be
calculated as the measured flux (listed in column~5) devided by $f_{\rm c}$.
The resulting value should,
however, be treated as an upper limit to the echo flux. The missing
pixels do not result from the field stars only, but also from the fact that
there were regions of the echo, not contaminated by stars, where the echo
was physically too faint to be measured.
Therefore, as a finally corrected flux, we took a mean value from the flux 
estimated in the above way and the measured flux (column~5). For all except the
observations from December~17, both fluxes were taken with the same weight
when averaging. The echo images obtained on December~17 show empty regions
significantly more extended than on earlier dates, so we took the measured
flux with a double weight when calculating the finally corrected flux for this
date. The echo fluxes corrected with the above procedure and converted to 
the ST magnitudes \citep{sirianni} are given in column~(7)
of Table~\ref{measur_tab}. The corrections to the echo total flux were
largest for the F435W filter, but even in this case they do not
exceed 10\% (0.10 mag.).

\begin{table*}\begin{minipage}[t]{\hsize} % 
\caption{Measurements of V838 Mon's light echo on images obtained
with HST/ACS.}
\label{measur_tab}
\centering\renewcommand{\footnoterule}{} % to avoid a line before footnotes
\begin{tabular}{cccccccc}
\hline
date & obs.mode & total & pixels & flux 
& surf.brightness & STmag & surf.brightness($f_{\rm c}=0.5$) \\
2002 &  & count s$^{-1}$ & number  
& W\,m$^{-2}$\,$\mu$m$^{-1}$ 
& W\,m$^{-2}$\,$\mu$m$^{-1}$\,arcsec$^{-2}$ & corrected
& W\,m$^{-2}$\,$\mu$m$^{-1}$\,arcsec$^{-2}$\\
 (1) & (2) & (3) & (4) & (5) & (6) & (7) & (8) \\
\hline
 & \\
 & \bf{F435W}\\
\hline
Apr.30 & polarimetric & 9.59E4 & 4.20E5 & 2.95E--13\,$\pm$7.3\% &
2.81E--16\,$\pm$10.7\% & 12.72\,$\pm$0.07 & 4.14E--16\,$\pm$7.5\%\\
May 20 & polarimetric & 7.55E4 & 4.80E5 & 2.33E--13\,$\pm$8.6\% &
1.94E--16\,$\pm$13.0\% & 12.92\,$\pm$0.08 & 2.65E--16\,$\pm$7.4\%\\
Sept.2 & standard & 3.58E4 & 9.44E5 & 1.10E--13\,$\pm$7.8\% & 
4.67E--17\,$\pm$12.0\% & 13.69\,$\pm$0.08 & 6.36E--17\,$\pm$7.5\%\\
Oct.28 & standard & 2.90E4 & 1.21E6 & 8.92E--14\,$\pm$7.3\% &
2.96E--17\,$\pm$10.3\% & 13.94\,$\pm$0.07 & 4.16E--17\,$\pm$7.5\%\\
Dec.17 & standard & 2.48E4 & 1.27E6 & 7.64E--14\,$\pm$11.2\% &
2.41E--17\,$\pm$17.5\% & 14.10\,$\pm$0.11 & 3.10E--17\,$\pm$7.4\%\\
\hline
 &\\
 &  \bf{F606W}\\
\hline
May 20 & polarimetric & 1.02E5 & 6.28E5 & 7.90E--13\,$\pm$5.0\% &
5.82E--16\,$\pm$7.5\% & 11.66\,$\pm$0.05 & 9.46E--16\,$\pm$5.8\%\\
Sept.2 & polarimetric & 5.19E4 & 1.16E6 & 4.01E--13\,$\pm$5.0\% &
1.41E--16\,$\pm$7.3\% & 12.39\,$\pm$0.06 & 2.25E--16\,$\pm$5.8\%\\
Oct.28 & standard & 3.94E4 & 1.36E6 & 3.05E--13\,$\pm$5.0\% &
8.93E--17\,$\pm$7.0\% & 12.68\,$\pm$0.05 & 1.44E--16\,$\pm$5.8\%\\
Dec.17 & polarimetric & 3.52E4 & 1.45E6 & 2.72E--13\,$\pm$5.0\% &
7.48E--17\,$\pm$7.2\% & 12.77\,$\pm$0.05 & 1.09E--16\,$\pm$5.7\%\\
\hline
 &\\
 &  \bf{F814W}\\
\hline
May 20 & standard & 1.29E5 & 5.07E5 & 8.98E--13\,$\pm$4.4\% &
7.08E--16\,$\pm$8.9\% & 11.49\,$\pm$0.05 & 1.16E--15\,$\pm$5.6\%\\
Sept.2 & standard & 7.04E4 & 1.07E6 & 4.89E--13\,$\pm$4.3\% &
1.83E--16\,$\pm$6.8\% & 12.14\,$\pm$0.05 & 2.87E--16\,$\pm$5.3\%\\
Oct.28 & standard & 5.96E4 & 1.36E6 & 4.14E--13\,$\pm$4.3\% &
1.22E--16\,$\pm$5.8\% & 12.34\,$\pm$0.04 & 1.97E--16\,$\pm$5.2\%\\
Dec.17 & standard & 5.40E4 & 1.51E6 & 3.75E--13\,$\pm$4.5\% &
9.95E--17\,$\pm$8.4\% & 12.43\,$\pm$0.05 & 1.50E--16\,$\pm$5.1\%\\
\hline\end{tabular}
\end{minipage}\end{table*}

As noted above, the covering factor, $f_{\rm c}$, is generally $<$1.0, but 
it varies from image to image. 
It is usually lowest for the F435W filter and shows a tendency to decrease with the
time of observations. As a result the surface brightness listed in
column~(6) of Table~\ref{measur_tab} is not entirely comparable between
different filters and different dates. This may create uncertainties and 
ambiguities when comparing the results of the echo simulations to 
the observational measurements. 
Therefore we also derived another mean surface brightness, which was calculated
from the brightest pixels in each image, counted until they fill $f_{\rm c} = 0.5$. 
The resulting values, named $S_{\rm B\,0.5}$ below, 
are listed in column~(8) of Table~\ref{measur_tab}.

Table~\ref{measur_tab} lists uncertainties of the obtained values
in addition to the results of measurements. The errors of the total flux and the
surface brightness (columns~5, 6, and 8) are expressed in percent of the values, 
while those of STmag (column~7) are given in magnitudes. When evaluating the
errors we took into account count statistics, uncertainities due to
subtracting the sky background, uncertainties of the echo dimension, and
calibration errors. The first source of errors, which we evaluated from
the standard deviation of the mean sky background, was negligible compared
to the other ones and resulted in errors $\la$0.1\% in the flux and surface
brightness, and $\la$0.001~mag in STmag. As noted in Sect.~\ref{obs_sect},
before measuring the echo, we
clipped the pixels with signal below 1.5$\sigma$ of the sky variations. This value
is of course somewhat arbitrary. To estimate the uncertainties caused by this
procedure we repeated measurments by allowing for 1$\sigma$ variations of
the clipping level. The surface brightness, $S_{\rm B}$, was most
vulnerable to this uncertainty, which was 4--7\% in the F606W and
F814W filters, and 8--16\% in the F435W filter. The total flux was
uncertain to 0.3--1.3\% in the F606W and F814W filters and 2--8\% in the
F435W filter. $S_{\rm B\,0.5}$ is insensitive to the
clipping level. Errors in determining the echo radius derived in
\citet{tss05} propagate through the correction procedure (described above) 
when deriving STmag, resulting in a $\sim$0.05~mag uncertainty. 
They also affect $S_{\rm B\,0.5}$ through the evaluation of the
total number of pixels inside the echo and result in
2.5--3.5\% uncertainty. The deriviation of $S_{\rm B}$
does not involve the echo radius. To evaluate uncertainties of the image flux
calibration, we selected ten field stars located outside
the light echoes and measured their corresponding counts in all images.
Calculating mean values and standard deviations,
we found that the calibration is uncertain to 7\%,
5\%, and 4\% in the F435W, F606W, and F814W filters, respectively.
Propagation of all these errors through the measuring procedure results in
the final errors listed in Table~\ref{measur_tab}.

%+++++++++++++++++++++++++++++++++++++++++++++++++++++++++++++++++++
\section{The model  \label{model_sect}}

Basic assumptions and formulae of the light echo geometry and 
its modelling can be found in \citet{tyl04} \citep[see also][]{sparks}.
Briefly, we assume a single-scattering approximation and that the echo
structure is described by the paraboloid equation, i.e.
\begin{equation}
  r = z + ct,
\label{parabol}
\end{equation}
where $r$ is a radial distance of the scattering point from the source of
radiation, $z$ is a projection of $r$ on the line of sight of the source, 
and $t$ is a time delay between the moment of observations of the scattered radiation 
and that of the radiation directly recorded from the source.

\begin{figure}
\includegraphics[scale=0.37]{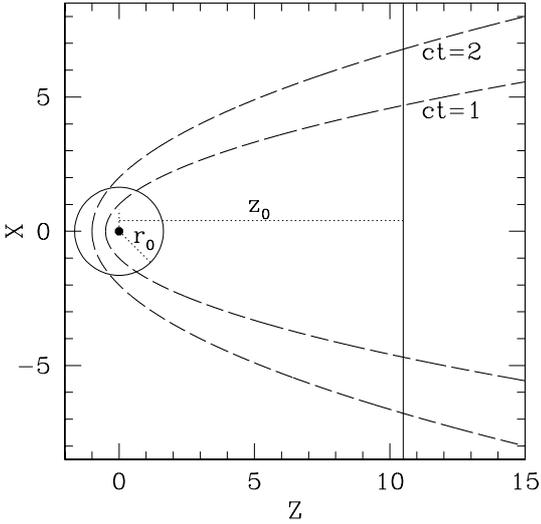}
\caption{Cross section, $y=0$, of the geometrical model adopted in 
our light-echo modelling. The source of
the light is at $x=0$, $y=0$, $z=0$ (black point in the figure). 
The observer is at $x=0$, $y=0$, $z=\infty$. 
$x$ and $z$ are in units of $ct$. Dust is uniformely
distributed in the space satisfying $z \le z_0$ and 
$r = \sqrt{x^2 + y^2 + z^2} \ge r_0$.
Dashed curves: light-echo paraboloids, Eq.~(\ref{parabol}), 
for two epochs differing by a factor of 2. If the source produced a
flash of light, which is observed between these two epochs, dust scattering the light 
at present ($ct=0$) towards 
the observer is situated between these two paraboloids.
}
\label{scatch_fig}
\end{figure}

The geometry assumed in our light-echo simulations is shown in
Fig.~\ref{scatch_fig}.
The echoing medium surrounding the flaring object is uniform and has 
a semi-infinite geometry. Its boundary is in a form of a plane
perpendicular to the line of sight, situated in front of the flaring object
at a certain distance, $z_0$, from the object. The medium extends well beyond the
object so that the light echo paraboloid never reaches its possible boundary
behind the object during the time span covered in our simulations. It was shown in
\citet{tyl04} that V838~Mon is situated in a cavity in the
surrounding dusty medium. We assume, for simplicity, that this dust-free hole is
spherical in respect to the object and has a radius $r_0$. 

In the above
approximation, the echoing medium is parametrized, apart from $z_0$ and
$r_0$, with the optical thickness of the medium along the line of sight of
the central object, $\tau_0$. The thickness devided by $z_0 - r_0$ gives 
the extinction coefficient, $Q_{\rm ext} = Q_{\rm abs} + Q_{\rm sca}$, where
$Q_{\rm abs}$ and $Q_{\rm sca}$ are the absorption and scattering coefficients,
respectively. The scattering on dust grains is usually anisotropic, so in
order to model it properly, we adopted a standard anisotropy phase function, i.e.
\begin{equation}
  f(\theta) = \frac{1 - g^2}{(1 + g^2 -2g\cos\theta)^{3/2}},
\end{equation}
where $\theta$ is a scattering angle and $g \equiv \langle \cos \theta \rangle$
is an anisotropy factor.
Generally, the anisotropy factor is
determined by properties of dust grains \citep[see e.g.][]{draine}. In our
modelling, $\langle \cos \theta \rangle$ is treated as a free parameter to
be determined from fitting the results of simulations to the observational
measurements.
Extinction of the light travelling through the echoing medium is taken into
account in the modelling.

The observed light curve of V838 Mon was obtained in the Johnson-Cousins
filters. The HST images of the light echo were obtained with the
HST filters, however, which differ from the Johnson-Cousins system. For the purpose
of our modelling we converted the photometric measurements of V838~Mon in the
$B$, $V$, $R_c$, and $I_c$, taken from the same references as in
\citet{tyl05}, into the light curves in F435W, F606W, and F814W
using transformation formulae of \citet{sirianni}. These light curves were
then used to model the evolution of the light echo images.

As discussed in Sect.~\ref{obs_sect}, the observed echo image was
contaminated by field stars, in particular by the diffraction pattern of
V838~Mon itself. The latter had a typical dimension of $\sim$3~arcsec in
radius (note that the diffraction cross was much more extended), 
and was removed in the
reduction precedure of all images before measuring the echo. 
This effect can be easily
taken into account in the model simulations and we removed the central
circular region of the above radius from calculating the model echo flux and
surface brightness.

\begin{figure}
\includegraphics[scale=0.37]{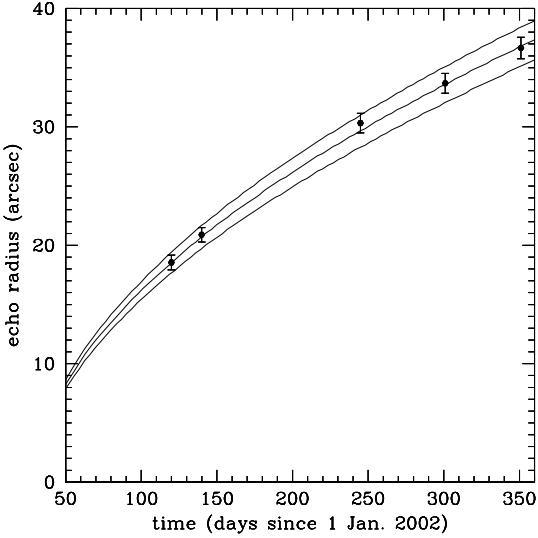}
\caption{Observed expansion of the outer rim of the light echo, as
measured in \citet{tss05} (symbols), compared to our modelling (lines) assuming a
distance to V838~Mon of 6.1~kpc. The curves were obtained taking $z_0$ =
2.1~pc (middle curve) $\pm$ 10\% (outer curves). The outer rim in the
simulations was defined as the radius at which the echo surface
brightness drops to 10\% of its mean value. The light curve of
V838~Mon, used in the simulations, is the same as that used to
model the light echo in the F435W filter, see upper panel of Fig.~\ref{mag_fig}.}
\label{radius_fig}
\end{figure}

Following \citet{sparks} and \citet{afbond}, we assume that V838~Mon is at a
distance of 6.1~kpc. For a fixed distance, the expansion 
of the echo outer rim is uniquely determined by $z_0$. The observed expansion of 
the light echo outer rim, as measured in \citet{tss05} and compared to the
results of our modelling in Fig.~\ref{radius_fig}, results in 
$z_0 \simeq 2.1$~pc. Following \citet{tyl04} and the adopted distance, we
assume $r_0 = 0.1$~pc. Note that the latter value is not crucial for 
the results of our analysis. The albedo of dust grains, 
i.e. $Q_{\rm sca}/Q_{\rm ext}$, was taken from
the interstellar extinction curve of \citet{wd01} with $R_V = 3.1$.

\section{Results of model fitting to observations  \label{res_sect}}

\begin{figure}
\includegraphics[scale=0.37]{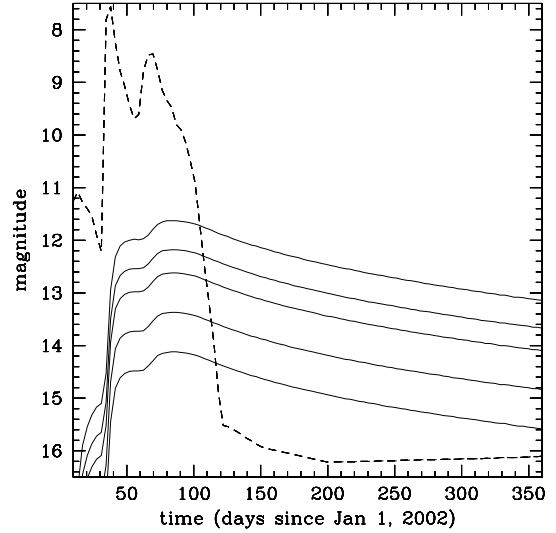}
\includegraphics[scale=0.37]{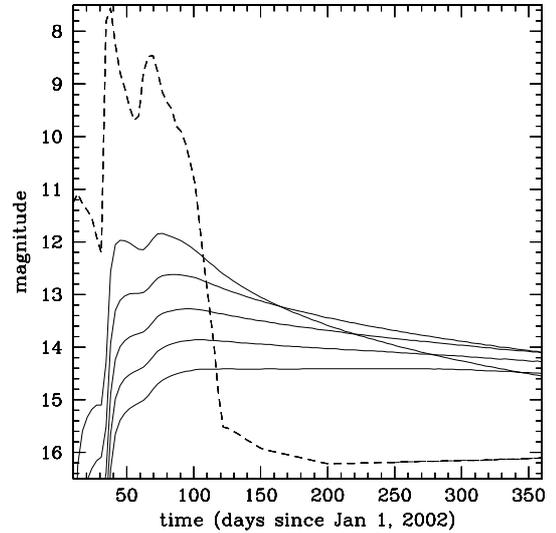}
\caption{Evolution of the model echo total flux (in the magnitude scale) 
as a function of the optical thickness, $\tau_0$, along the line of sight of
the central source (upper panel) and the anisotropy factor, 
$\langle \cos \theta \rangle$, (bottom panel). 
Dashed curve in both panels: the light curve of the central source (the
same as in the upper panel of Fig.~\ref{mag_fig}). Full curves in the upper panel: 
the light echo flux evolution with $\langle \cos \theta \rangle$ fixed at 0.6
but for $\tau_0$ varying from 0.05 (bottom curve), through 0.1, 0.2, 0.3, and 0.5 
(uppermost curve). 
Full curves in the bottom panel: the light echo flux evolution with 
$\tau_0$ fixed at 0.2 but for $\langle \cos \theta \rangle$ varying from
0.0 (bottom curve), through 0.2, 0.4, 0.6, and 0.8 (uppermost curve)}
\label{taucos_fig}
\end{figure}

The results of our modelling with the fixed values of the distance, 
$z_0$, $r_0$, and
albedo, as described in Sect.~\ref{model_sect}, depend on
two parameters, i.e. optical thickness of the echoing matter along the
line of sight in front of V838~Mon, $\tau_0$, and the scattering anisotropy
factor, $\langle \cos \theta \rangle$. This is illustrated in
Fig.~\ref{taucos_fig}.
Note that each of the two parameters affects the model results
in a different way. As can be seen from the upper panel of Fig.~\ref{taucos_fig},
$\tau_0$ does not change the shape of the echo flux evolution but it moves
the curve vertically in the diagram.
On the other hand, $\langle \cos \theta \rangle$ 
determines the rate of fading of the light echo with time 
(see the bottom panel of Fig.~\ref{taucos_fig}). 
Therefore fitting the model results
to the observed evolution of the light echo in different filters can
constrain the values of these two parameters and their dependence on the
wavelength. 

\begin{figure}
\includegraphics[scale=0.37]{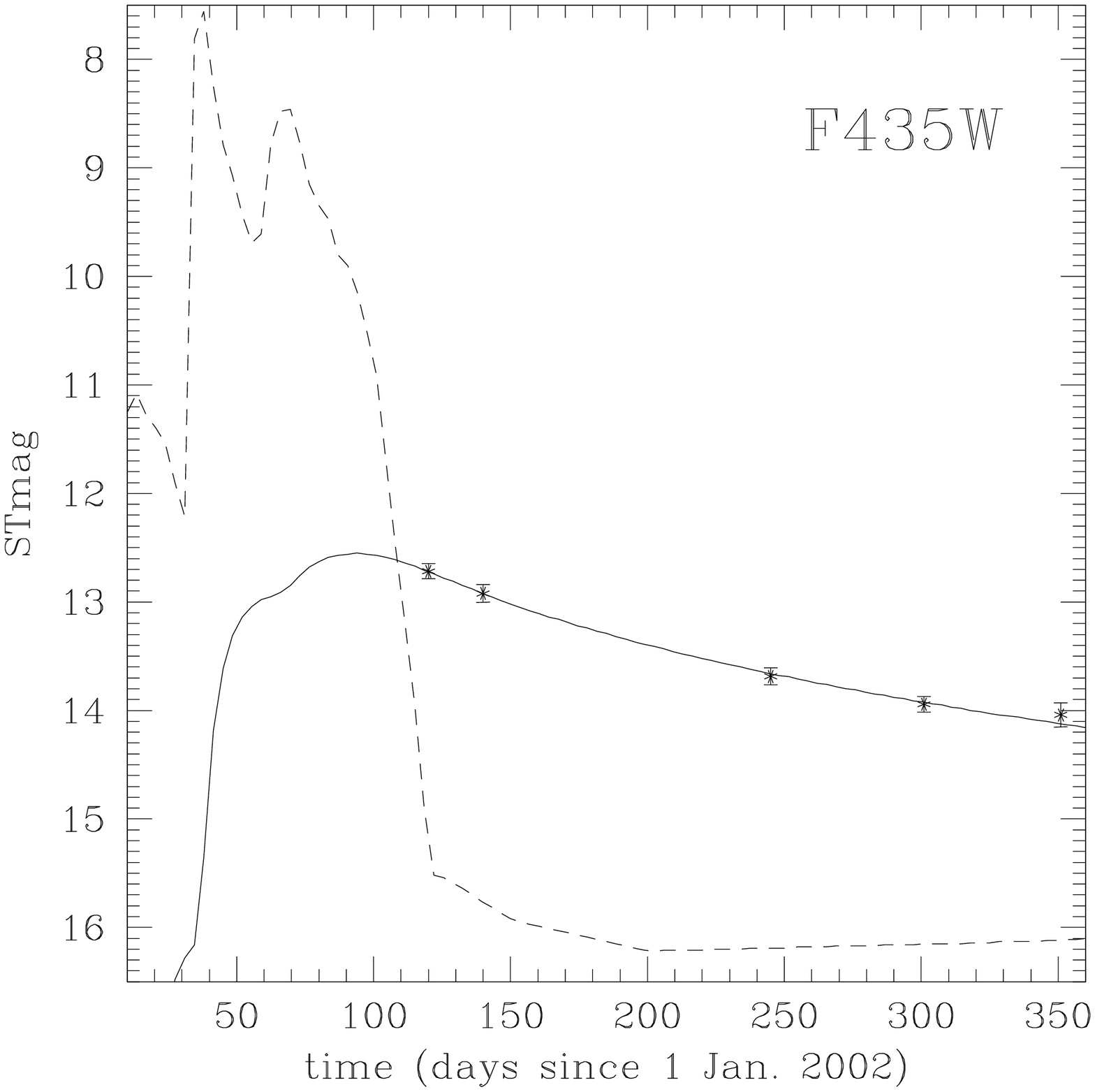}
\includegraphics[scale=0.37]{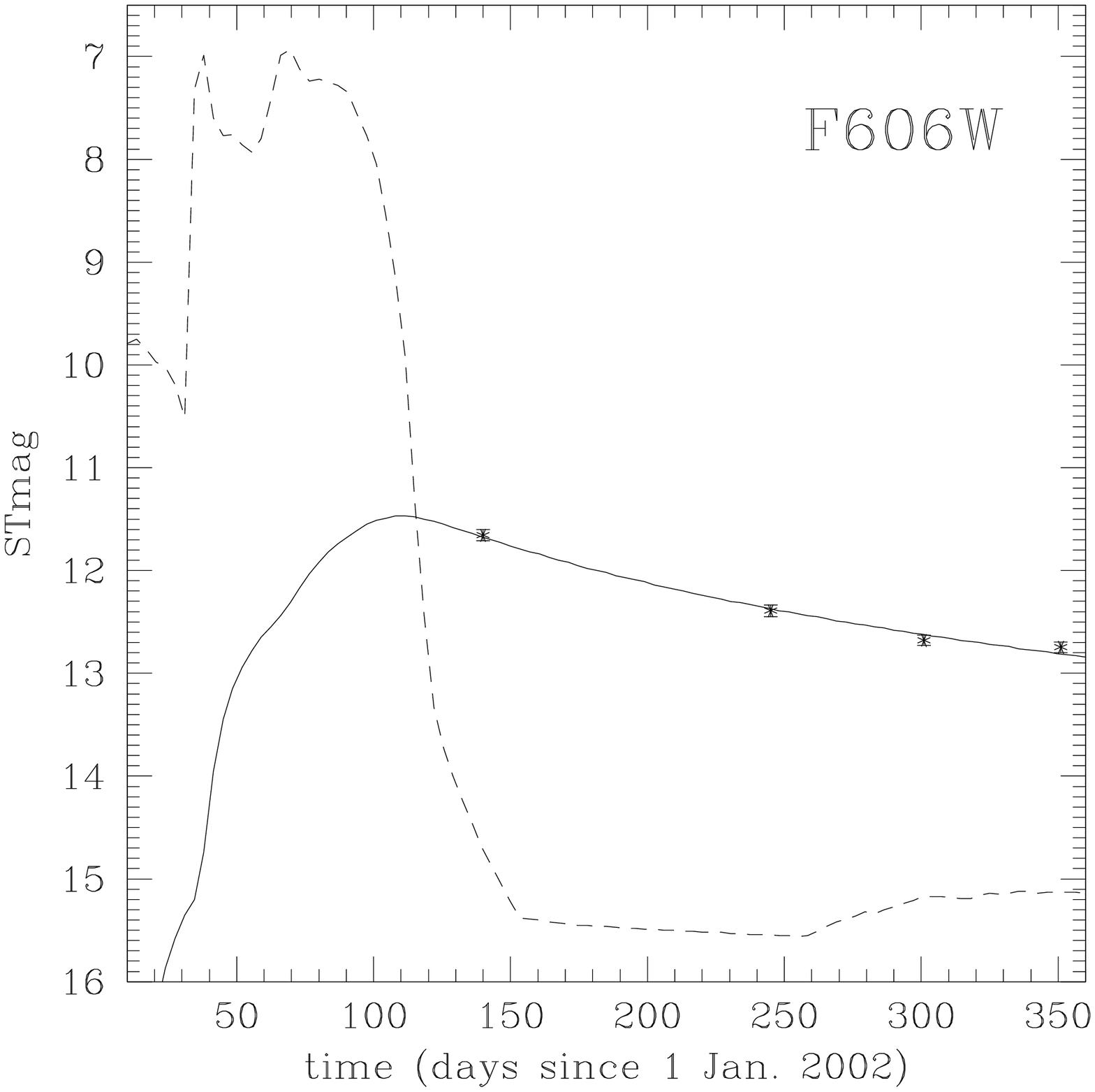}
\includegraphics[scale=0.37]{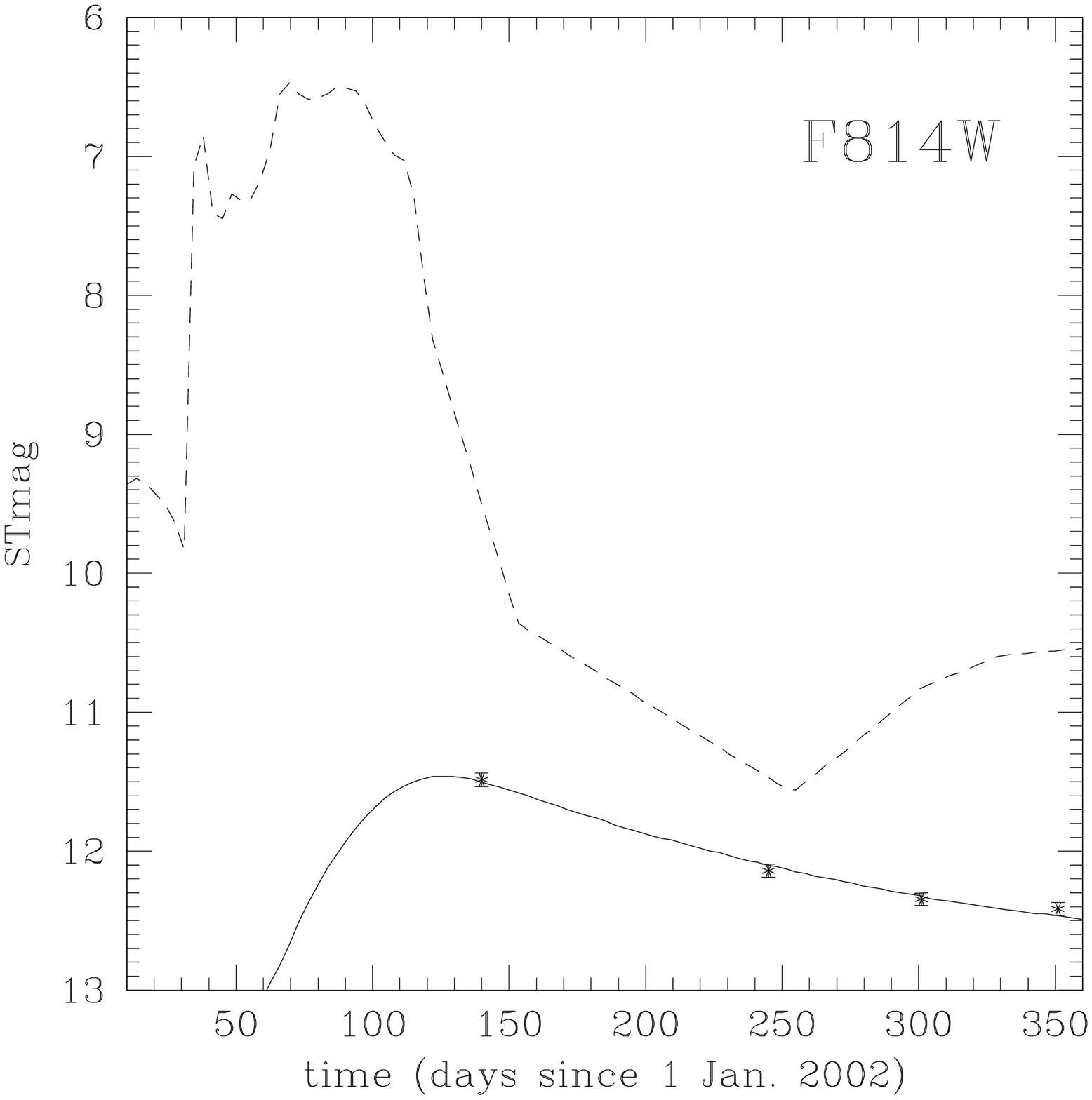}
\caption{Evolution of the light-echo total-flux in the
ST magnitude scale as observed in the F435W (upper panel), F606W (middle
panel), and F814W (bottom panel) filters. 
Dashed curve: observed light curve of V838 Mon. Asterisks: 
observed light-echo magnitudes (column~7 in Table~\ref{measur_tab}). 
Full curve: best fit of the modelled evolution to the observational points.
Parameters of the fit can be found in Table~\ref{res_tab} upper rows. }
\label{mag_fig}
\end{figure}

\begin{figure}
\includegraphics[scale=0.37]{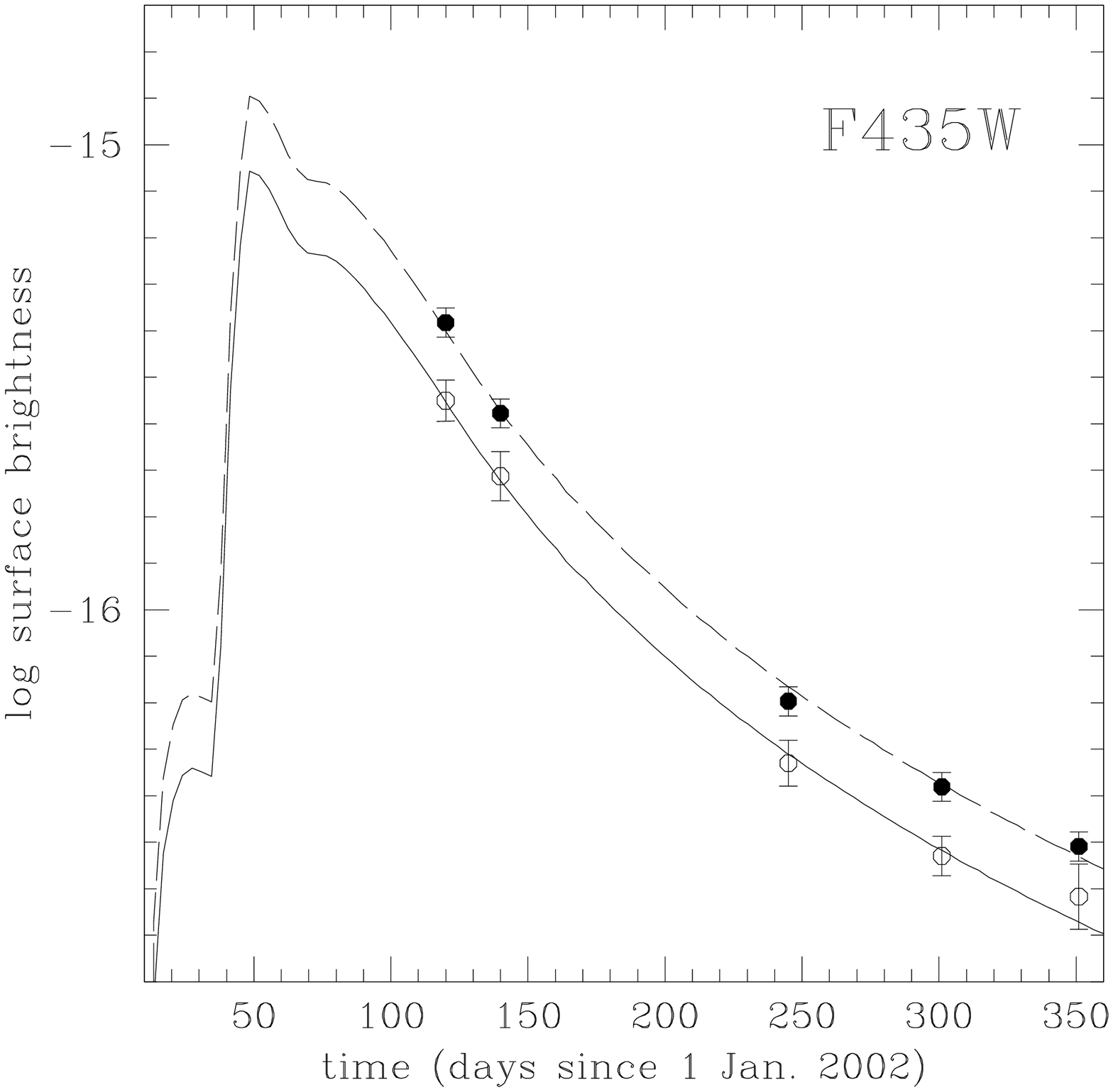}
\includegraphics[scale=0.37]{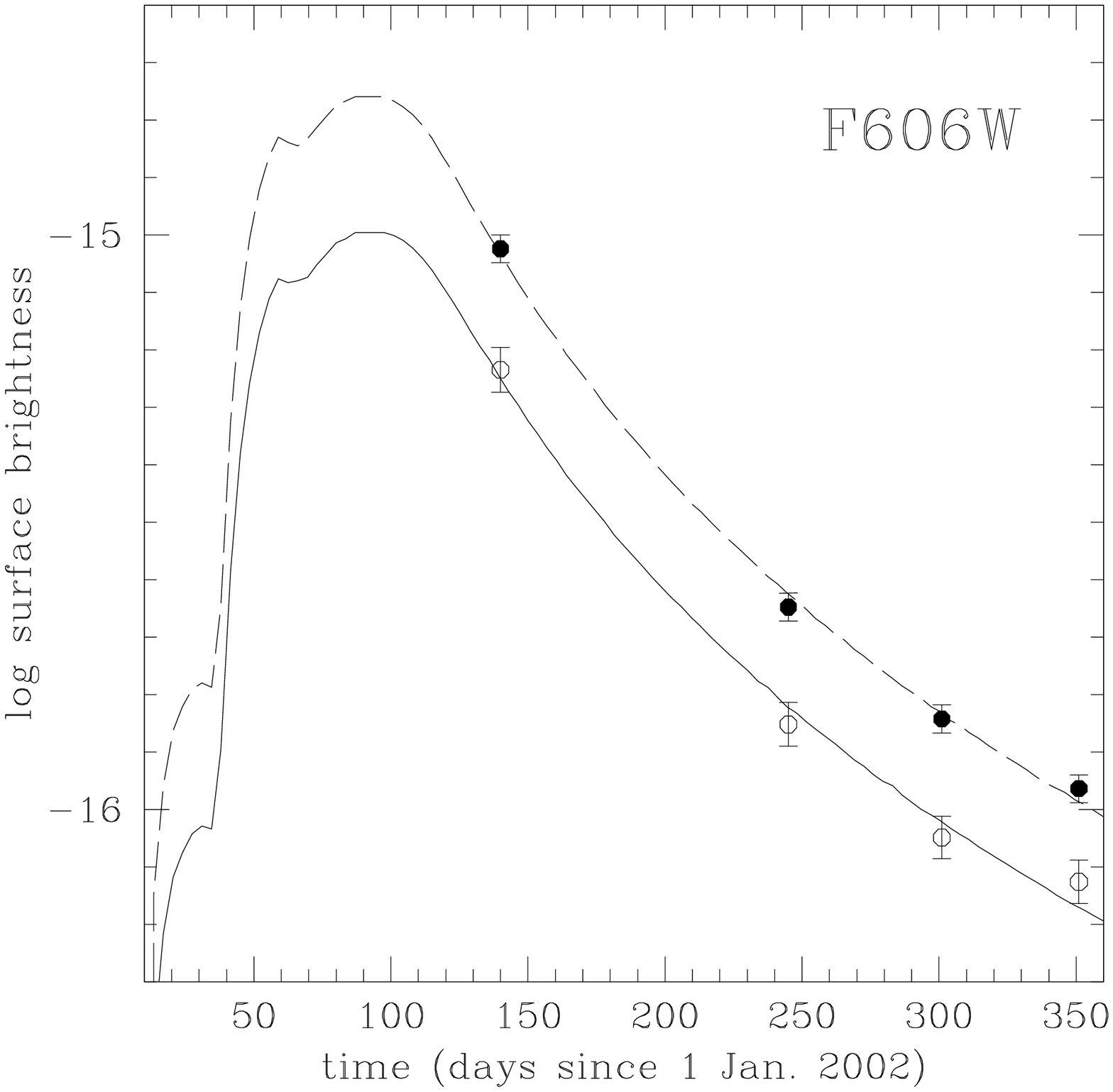}
\includegraphics[scale=0.37]{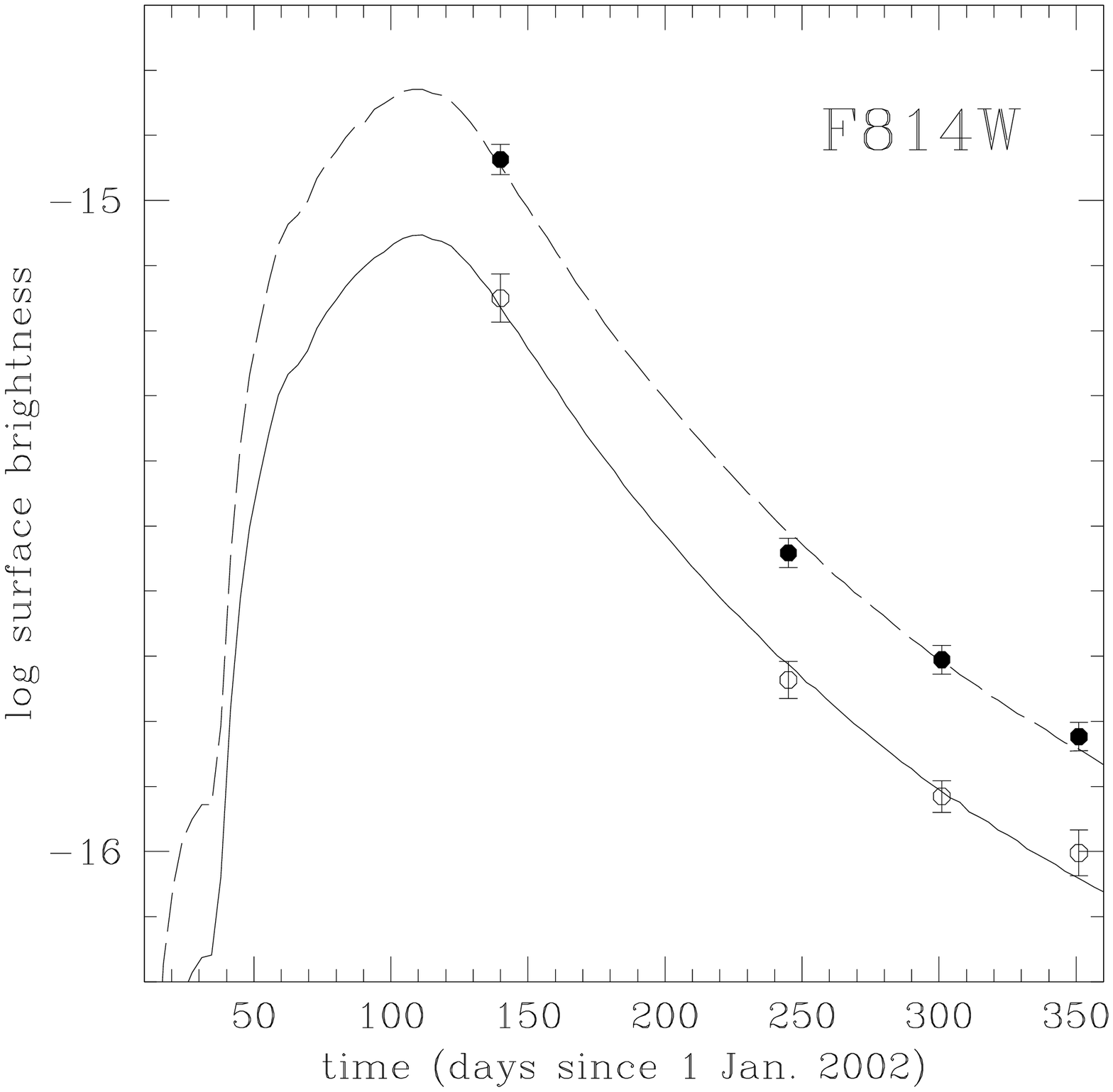}
\caption{Evolution of the light-echo surface-brightness in the F435W (upper
panel), F606W (middle panel), and F814W (bottom panel) filters. 
Open symbols and full curve:
observed light-echo surface-brightness $S_{\rm B}$
(column~6 in Table~\ref{measur_tab}) and its best-fitting model.
Filled symbols and dashed curve: the same but for $S_{\rm B\,0.5}$
(column~8 in Table~\ref{measur_tab}).
Parameters of the fit can be found in Table~\ref{res_tab} middle and
bottom rows.}
\label{sb_fig}
\end{figure}

Figures \ref{mag_fig} and \ref{sb_fig} compare the results of our models of the light
echo (curves) fitted to the observational measurements
(symbols with error bars) collected in Table~\ref{measur_tab}.
The fitting was made with the $\chi^2$ method. 
Figure~\ref{mag_fig}
compares the total brightness of the light echo in the
ST magnitude scale (data from column 7 in Table~\ref{measur_tab}). 
Here, we also plotted the
observed light curves of V838~Mon (dashed curves) that were used
to model the echo.
Figure~\ref{sb_fig} presents the observed and model evolution
of the mean surface brightness of the echo. Open symbols and full curves
refer to the measurments and modelling of $S_{\rm B}$ (data from column 6 in
Table~\ref{measur_tab}). Filled symbols and dashed curves shows the same but
for $S_{\rm B\,0.5}$ (data from column 8 in Table~\ref{measur_tab}).
The parameters of all models displayed in 
Figs.~\ref{mag_fig} and \ref{sb_fig} are given in Table~\ref{res_tab}.
The errors of the parameters were obtained form the 90\% confidence level of
the $\chi^2$ fitting.

\begin{table}
\caption{Values of $\tau_0$ and $\langle \cos \theta \rangle$ derived from the
model fitting to the observations.}
\label{res_tab}
\begin{tabular}{cccc}
\hline
~~~~~~~~~~ & ~~~~F435W~~~~ & ~~~~F606W~~~~ & ~~~~F814W~~~~ \\
\hline
\end{tabular}
  \\
 total flux (STmag) \\
\begin{tabular}{cccc}
 $\tau_0$ & $0.204 \pm .016$ &  $0.151 \pm .003$ &  $0.087 \pm .003$ \\
 $\langle \cos \theta \rangle$ & $0.67 \pm .06$ & $0.62 \pm .02$ & $0.59 \pm .03$ \\
%\hline
\end{tabular}
 \\
 \\
 surface brightness $S_{\rm B}$ \\
\begin{tabular}{cccc}
 $\tau_0$ & $0.218 \pm .028$ & $0.159 \pm .009$ & $0.087 \pm .006$ \\
 $\langle \cos \theta \rangle$ & $0.64 \pm .10$ & $0.57 \pm .06$ & $0.54 \pm .06$ \\
\end{tabular}
 \\
 \\
 surface brightness $S_{\rm B\,0.5}$ \\
\begin{tabular}{cccc}
 $\tau_0$ & $0.306 \pm .020$ & $0.249 \pm .010$ & $0.137 \pm .004$ \\
 $\langle \cos \theta \rangle$ & $0.64 \pm .05$ & $0.60 \pm .04$ & $0.56 \pm .03$ \\
\hline
\end{tabular}
\end{table}

\section{Analysis and discussion  \label{discuss_sect}}

\begin{figure}
\includegraphics[scale=0.37]{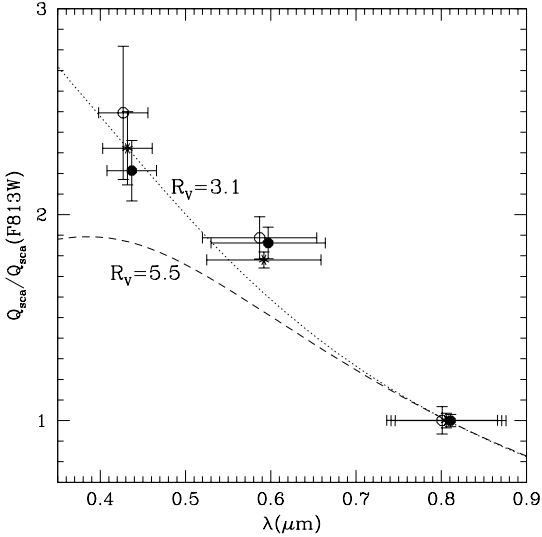}
\caption{Dependence of the scattering coefficient, $Q_{\rm sca}$, 
on the wavelength as derived from modelling of the echo. The values are
normalized to $Q_{\rm sca}$ at the effective wavelength of the F814W filter. 
Asterisks: results of 
the models fitting the total flux, STmag, 
(upper rows in Table~\ref{res_tab}). Open circles: results of 
the models fitting the mean surface brightness, $S_{\rm b}$, 
(middle rows in Table~\ref{res_tab}).
Filled circles: results of 
the models fitting the mean surface brightness derived from the brightest
pixels filling $f_{\rm c} = 0.5$, $S_{\rm B\,0.5}$,
(bottom rows in Table~\ref{res_tab}).
The horizontal error bars represent the widths of the photometric bands.
Open and filled symbols are
shifted by $\pm 0.05\,\mu$m in the abscissa axis for clarity.
Curves: relations expected from modelling the interstellar extinction curve
of \citet{wd01} for $R_{V} = 3.1$ (dotted) and $R_{V} = 5.5$
(dashed).}
\label{qsca_fig}
\end{figure}
 
\begin{figure}
\includegraphics[scale=0.37]{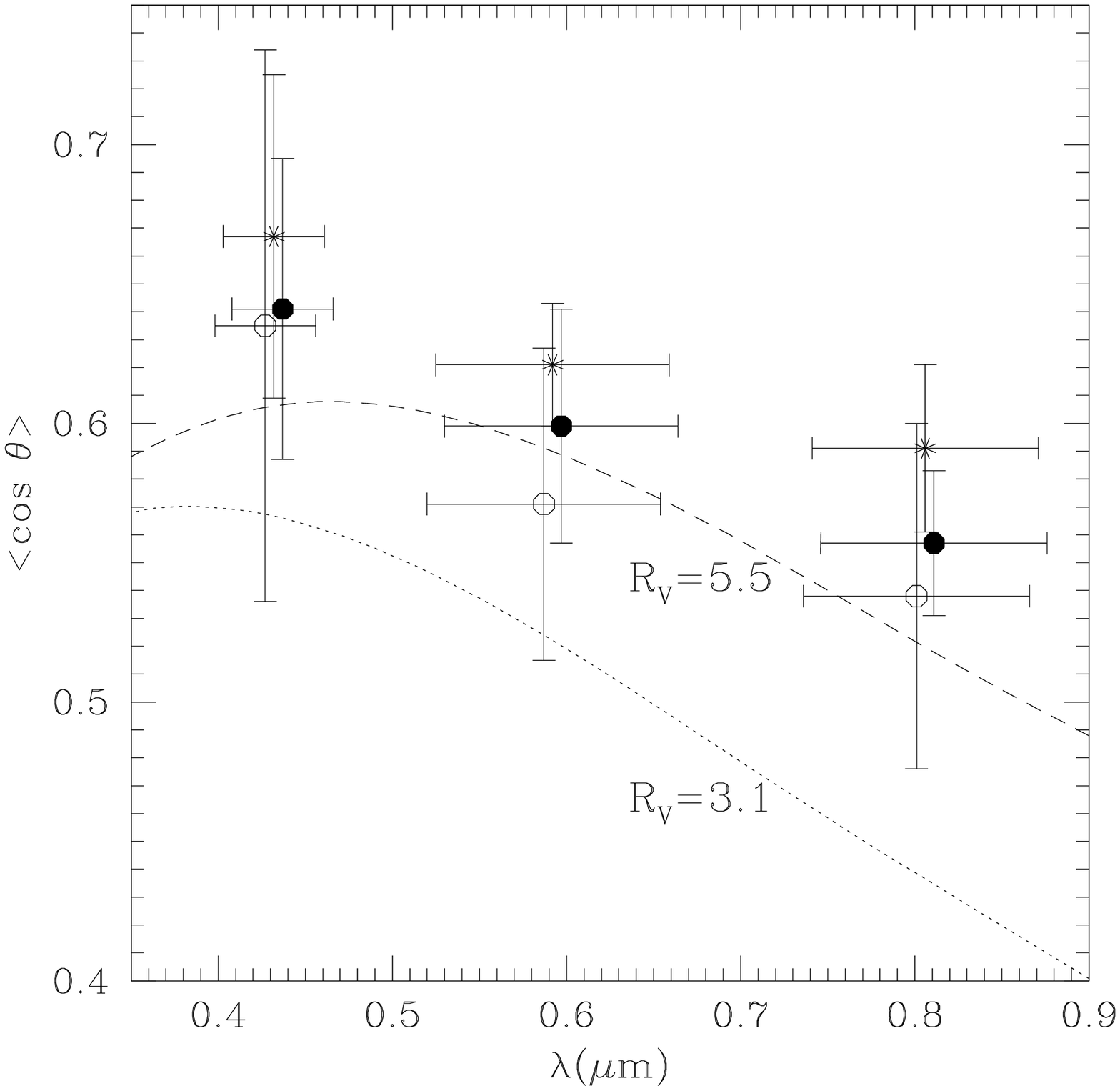}
\caption{Dependence of the anisotropy factor, $\langle \cos \theta \rangle$,
on the wavelength, as derived from our modelling of the echo. 
The symbols and curves have the same meaning as in Fig.~\ref{qsca_fig}.}
\label{cos_fig}
\end{figure}
 
As can be seen from Table \ref{res_tab}, the optical thickness of the dusty
matter in front of V838~Mon, $\tau_0$, as determined from modelling the surface
brightness, $S_{\rm B}$, (middle rows in the table) is systematically higher
than the values derived from fitting the total echo flux (upper rows in the
table). The difference is within the error bars, however.
Modelling of the surface brightness of the
bright regions covering 50\% of the echo surface, $S_{\rm B\,0.5}$, 
(bottom rows in the table) gives significantly
higher values of $\tau_0$ than in the two previous cases. This is not
surprising, as brighter regions mean more effective scattering and thus
a higher extinction coefficient. The values listed in the upper and middle
rows of Table~\ref{res_tab} can be considered as typical for the echoing
medium of V838~Mon. Those in the bottom rows, when compared to the other
ones, illustrate how inhomogeneous the medium is.

The results in Table~\ref{res_tab} allow us to study properties of dust
grains in the echoing medium. Figures~\ref{qsca_fig} and \ref{cos_fig} show
values of the scattering coefficient and the anisotropy factor,
respectively, as
functions of wavelength and compare them to the analogous values obtained by
\citet{wd01} from modelling the interstellar extinction curve with 
$R_{V} = 3.1$ and 5.5.

As can be seen from Fig.~\ref{qsca_fig}, our modelling of different
observational quantities (total flux, surface brightness)
results in a similar wavelength dependence of the
scattering coefficient.
We can also conclude that this relation
is not significantly different from that expected from the standard composition 
and size
distribution of dust grains used to model the interstellar extinction curve
of \citet{wd01} with $R_{V} = 3.1$. This suggests that the dusty matter 
responsible for the observed echo is not significantly different from 
the typical interstellar dusty medium.

The scattering anisotropy factor, $\langle \cos \theta \rangle$, as
displayed in Fig.~\ref{cos_fig}, is more sensitive to the observational
uncertainties than the scattering coefficient. There are systematic
differences in the derived $\langle \cos \theta \rangle$ values, depending
on which quantity was analysed, but the differences are
within the error bars.
The value of $\langle \cos \theta \rangle$ is determined
from the rate of the echo fading with time. This rate, if derived from the
measurements of all pixels detecting emission from the echo, is likely to be
distorted because the portion of the echo surface covered by these
pixels is different for different dates of the observations. Therefore we expect
that the results of fitting the mean surface brightness, $S_{\rm B\,0.5}$,
(bottom row in
Table~\ref{res_tab} and filled circles in Fig.~\ref{cos_fig}) are more
reliable than the other two. 

Fig.~\ref{cos_fig} shows that our
determinations of $\langle \cos \theta \rangle$ gave values
systematically greater than those expected from
the interestellar extinction curve with $R_{V} = 3.1$. They are better
reproduced by the extinction curve with $R_{V} = 5.5$, but even in this
case the points  mostly lie above the curve. 
Greater $\langle \cos \theta \rangle$
is expected for bigger dust grains, but this usually implies
that the scattering coefficient is less dependent on the wavelength, 
which would be in
conflict with the points in Fig.~\ref{qsca_fig}. However, the model
extinction curve and $\langle \cos \theta \rangle$ depend not only on the
grain size, but also on the chemical composition and structure of the
grains. It is thus possible that playing with different parameters of the
dust grains would allow one to better reproduce the dust properties we
obtained from the analysis of the light echo of V838~Mon. This is
out of the scope of the present study, however.

We can estimate from Table~\ref{res_tab} (upper and middle rows) 
that the additional extinction 
of V838~Mon due to local dust responsible for the light echo is 
$A_{V} \simeq 1.2\,\tau_0({\rm F606W}) \simeq 0.18$ ($E_{B-V} \simeq
0.05$). Of course, this is the most likely value derived from the mean
properties of the local dusty medium seen in the echo. This medium is
inhomogeneous, however, as is shown by the uneven distribution of the surface
brightness over the echo images. Therefore the real extinction along the
line of sight of V838~Mon can be different from the above
estimate. Nevertheless, the results obtained from the brighter half part of the
echo surface, i.e. from  fitting $S_{\rm B\,0.5}$
(bottom rows in Table~\ref{res_tab}), indicate that it is
rather improbable that $A_{V} > 0.3$. With the observed extinction curve
(Fig.~\ref{qsca_fig}) this would also imply $E_{B-V} \ga 0.10$ from the
local dust, which would be too large compared to the observed difference in
$E_{B-V}$ between V838~Mon and the three B-type main-sequence stars observed 
by \citet{afbond} (see Sect.~\ref{intro}).

Using the standard interstellar-medium relation, 
$A_{V}/N_{\rm H} \simeq 5.3\times10^{-22}$~cm$^2$
\citep{bohlin}, where $N_{\rm H}$ is the column density of hydrogen
(atomic plus molecular), and taking
$A_{V} \simeq 0.18$, we obtain a surface density of the echoing matter 
in front of V838~Mon to be $\sim 8 \times 10^{-4}$~g\,cm$^{-2}$. 
In 2004 the angular radius of the light echo reached 
$\sim$1~arcmin \citep{tss05}. With a distance of 6.1~kpc this allows 
us to estimate
the mass of the diffuse matter of the light echo to be $\sim$35~M$_\odot$.
This is a lower limit to the mass of the circumstellar matter of V838~Mon,
since, first, in later epochs the echo expanded to even larger radii, second, the
estimate is based on the optical thickness {\it in front of} V838~Mon
only, while the matter almost certainly extends beyond the object. 
The above estimate can be compared to 90--100~M$_\odot$ obtained by
\citet{ktd} for the
molecular complex seen on the CO rotational lines in the vicinity of
V838~Mon and a hundred M$_\odot$ estimated by \citet{baner06} for the mass of the
echoing matter in the infrared. All these estimates clearly show that 
the diffuse matter in the vicinity of V828~Mon, partly seen in the light
echo, could not have resulted from a past mass loss activity of V838~Mon, as advocated,
e.g. in \citet{bond03} and \citet{bond07}. It is much more probable that
we see remnants of an interstellar complex, from which V838~Mon, its
companion, and perhaps the other members of the observed cluster were formed.

It is worth noting that \citet{kervella} recently found that the mass of
the dusty matter echoing pulsating radiation from the Cepheid RS~Pup is
$\sim$290~M$_\odot$. It seems therefore that, in general, a stellar light echo to be
well resolved in imaging observations requires a flaring star to be
associated with a dusty medium of interstellar origin rather than formed from a
mass loss activity of the star.

On the basis of our results, we can quite safely conclude that 
the optical thickness of 
the dusty matter seen in the light echo is too low to account for
the observed dimming of the B-type companion by $\sim$1.3~mag 
compared to the other B-type stars observed by \citet{afbond}
in the vicinity of V838~Mon (see Sect.~\ref{intro}). Other
explanations of this discrepancy have to be considered, e.g.
mis-classification of the B-type companion, an evolutionary status of 
the companion different from that of the three B-type stars of the cluster, 
or a mistaken distance to V838~Mon and its companion.

The classification of the companion as a B3 main-sequence star comes from
\citet[][see also \citet{mnv07}]{mdh02}. 
This classification was essentially confirmed by
\citet{afbond}, but the B-type component in their spectrum, similarly as in
the spectrum of \citet{kst09}, was then significantly contaminated by 
V838~Mon.
The observed $V$ brightness of the companion would be in accord with those of the
three B-type stars of \citet{afbond} if it were of a $\sim$B7\,V spectral type.
The intrinsic colours of such a star would be $(B-V)_0 = -0.13$ and $(U-B)_0
= -0.43$ \citep{sk82}. From Goranskij's$^1$ compilation of the photometric data
we derived $V = 16.19 \pm 0.03$, $B - V = 0.55 \pm 0.05$, and
$U - B = -0.15 \pm 0.07$ for the B-type companion. The observed $B-V$ colour 
can be reconciled with that of the above standard if $E_{B-V} \simeq 0.69$. But then
the reddened $U - B$ colour of the standard star is $+0.08$, i.e. it is
significantly redder than the observed one. Using the photometric
measurements of the B-type companion of \citet{mnv07}, 
we obtain $E_{B-V} \simeq 0.82$ and $U-B = +0.17$ for the reddened standard, 
which is to be compared to the observed one of $-0.06$. We
thus conclude that B7\,V cannot be reconciled with the observed
colours of the B-type companion. Note that the best fit of a standard star 
to the colours derived from the Goranskij data can be obtained taking 
the spectral type of B2.5\,V and $E_{B-V} \simeq 0.80$, which is not significantly
different from the B3\,V spectral type obtained by \citet{mdh02}.

The second possibility is that the B-type companion of
V838~Mon is intrinsically fainter because it is
in a somewhat different evolutionary stage than the three B-type stars of
\citet{afbond}.\footnote{As sugested by
V.~Goranskij during the STScI workshop "Intermediate-Luminosity Red
Transients", June 2011.} All existing observational data
indicate that all four stars are on the main sequence (MS) or not far
from it.
The only reasonable situation to be discussed in this context is that
the B-type companion of V838~Mon has a similar effective temperature but is 
less luminous (by $\sim$1.3 mag.) than star~9 (B3\,V) of \citet{afbond} 
because it is less massive but is
at its hottest stage, namely near the zero age main sequence (ZAMS), 
while star~9 is more evolved on the MS.
Taking the calibrations of the absolute magnitude versus spectral type 
and luminosity class from \citet{sk82},
one easily concludes that to account for 
a $\sim$1.3 magnitude difference between two B-type stars of a similar
spectral type with one of them being on ZAMS,
the second star has to be of luminosity class~IV, i.e.
slightly off MS. This would probably not be in a significant conflict 
with the spectroscopically determined 
class~V, given the quality of the data of \citet{afbond}. However, the
differences in magnitude between star~9, star~8 (B4\,V), and star~7 (B6\,V) 
imply that the latter two stars have also to be of luminosity class~IV. To
have a situation in which three stars of different masses are about to leave
or just left MS at the same time, one has to conclude that the three stars
have different ages, e.g. star~7 would have to be about twice as old as
star~9. This disproves the idea that these three stars form a cluster.
Futhermore, a case in which three B-type stars formed at different times
happen to lie in a close vicinity in the
sky and mimic an isochronic sequence of a
cluster in the colour-magnitude diagram is extremely improbable. So is
the whole scenario discussed in this paragraph.

If two MS stars of the same spectral type suffer from a similar extinction
but are significantly different in the observed brightness then the only
reasonable conclusion is that the stars are at different distances. In this
way we return to the study of \citet{muna07}, where the authors concluded,
primarily on the basis of the spectroscopic distance of the B-type
companion, that V838~Mon is at a distance of $\sim$10~kpc. In this case
V838~Mon and its companion would have nothing to do with the cluster of
\citet{afbond}. This possibility creates a problem with the observed
linear polarization of the light echo, however, as analysed in \citet{sparks}. 
To reconcile the
latter with the distance of 10~kpc, one has to postpone that maximum
polarization occurs at a scattering angle of $\sim$60\degr\ and not at
90\degr, as assumed in \citet{sparks}. Theoretically this cannot be excluded
\citep[e.g. Chapter 4.2 in][]{krug}
but requires rather unusual dust grains, e.g. metallic
particles.

We performed light echo simulations, similar to those described in
Sect.~\ref{res_sect}, but assuming a distance of 10~kpc. They gave results
not significantly different from those obtained in Sect.~\ref{res_sect},
except that
$\langle \cos \theta \rangle$ was systematically higher by $\sim$20\% 
than the values in Table~\ref{res_tab}. 

We can conclude that if the
distance of V838~Mon were $\sim$10~kpc, dust in the echoing medium would
have to be quite peculiar, 
i.e. giving an extinction curve close to the standard one
but scattering strongly in forward directions and 
giving maximum polarization at scattering angles significantly lower than
90\degr.
With the present state of our knowledge of interstellar dust grains, we
consider the above result as an argument against a distance as large as
$\sim$10~kpc.

In summary, we cannot present any strightforward
scenario that could explain the $\sim$1.3 magnitude difference between the
B-type companion of V838~Mon and the B3\,V star of \citet{afbond} and which
would be in accord 
with the observed properties of dust in the echoing matter of V838~Mon.
Perhaps this could be a dust cloud giving a grey extinction, lying well in
front of V838~Mon (not to be seen in the light echo), and not intervening
with the lines of sight of the three B-type stars of \citet{afbond}. A
detailed spectroscopic study, particularly of the interstellar
line profiles (Na\,I, K\,I), as well as multi-colour photometric measurements 
of a large sample of stars in the field of V838~Mon would perhaps enable one
to be conclusive in this subject.

\acknowledgements{The research
reported on in this paper has partly been supported by a grant no.
N\,N203\,403939 financed by the Polish Ministry of Sciences and Higher
Education. }
   
\bibliographystyle{aa}

\end{document}